\documentclass[aps,nofootinbib,showpacs,preprintnumbers,twocolumn,superscriptaddress,prl]{revtex4-1}

\usepackage{amsmath, amsthm, amssymb}
\usepackage{graphicx}
\usepackage{bm}
\usepackage{epsfig}
\usepackage[T1]{fontenc}
\usepackage[latin9]{inputenc}
\usepackage{color}
\usepackage{hyperref}
\usepackage{subfigure}

\begin{document}

\title{Odd-parity triplet superconductivity in multi-orbital materials with strong spin-orbit coupling: applications to doped
Sr$_{2}$IrO$_{4}$}
\author{Zi Yang Meng}
%\email{zymeng@physics.utoronto.ca}
\affiliation{Department of Physics, University of Toronto, Toronto, Ontario M5S 1A7, Canada}
\author{Yong Baek Kim}
\affiliation{Department of Physics, University of Toronto, Toronto, Ontario M5S 1A7, Canada}
\affiliation{Canadian Institute for Advanced Research/Quantum Materials Program, Toronto, Ontario MSG 1Z8, Canada}
\affiliation{School of Physics, Korea Institute for Advanced Study, Seoul 130-722, Korea}
\author{Hae-Young Kee}
\email{hykee@physics.utoronto.ca}
\affiliation{Department of Physics, University of Toronto, Toronto, Ontario M5S 1A7, Canada}
\affiliation{Canadian Institute for Advanced Research/Quantum Materials Program, Toronto, Ontario MSG 1Z8, Canada}
\date{\today}

\begin{abstract}
We explore possible superconducting states in $t_{2g}$ multi-orbital correlated electron systems with strong spin-orbit coupling (SOC). In order to study such systems in a controlled manner, we employ large-scale dynamical mean-field theory (DMFT) simulations with the hybridization expansion continuous-time Quantum Monte Carlo (CTQMC) impurity solver. To determine the pairing symmetry, we go beyond the local DMFT formalism using parquet equations to introduce the momentum dependence in the two-particle vertex and correlation functions.
In the strong SOC limit, a singlet, $d$-wave pairing state in the electron-doped side of the phase diagram is observed at weak Hund's coupling, which is triggered by antiferromagnetic fluctuations. When the Hund's coupling is comparable to SOC, a two-fold degenerate, triplet $p$-wave pairing state with relatively high transition temperature emerges in the hole-doped side of the phase diagram, which is associated with enhanced charge fluctuations. Experimental implications to doped Sr$_{2}$IrO$_{4}$ are discussed.
\end{abstract}

% http://www.aip.org/pacs/pacs2010/individuals/pacs2010_regular_edition
% 71.27.+a Strongly correlated electron systems
% 71.10.Fd Hubbard model electronic structure
% 74.70.-b Superconducting materials, noncuprate materials
% 74.70.Kn Superconducting materials, organic compounds
% 74.20.-z Superconductivity, theories and models of
% 74.20.Mn Superconductivity, nonconventional mechanism
% 74.20.Rp Superconductivity, pairing symmetries
% 71.30.+h Metal-insulator transition, 71.30.+h
% 73.21.Ac Multilayers, electron states and collective excitations in
% 75.70.Cn Multilayers, magnetic properties of

\maketitle

\paragraph*{Introduction.-}
The investigation of novel electronic states in correlated electron systems with spin-orbit coupling has been a recent subject of intensive research~\cite{Krempa14}. Early experiments that prompted such activities are the studies of
the iridium perovskite oxide Sr$_{2}$IrO$_{4}$~\cite{Cao98,BJKim08,BJKim09,Fujiyama12,Kim12,Carter12,Carter13,Ye13,Li13}. Due to strong SOC, the $t_{2g}$ orbitals of Ir$^{4+}$ ions split into $J_{\text{eff}}=1/2$ doublet and $J_{\text{eff}}=3/2$ quadruplet, leading to a spin-orbit-induced Mott insulator, with a moderate Hubbard interaction $U$.
%where the spin-orbit coupling splits the $t_{2g}$
%orbitals of Ir$^{4+}$ ions into $J_{\text{eff}}=1/2$ doublet and $J_{\text{eff}}=3/2$ quadruplet. It is suggested that the band-width of the 
%half-filled $J_{\text{eff}}=1/2$ band is small and a moderate strength of the Hubbard $U$ is enough to produce a spin-orbit-induced Mott insulator.  
Given the similarity in lattice structure and Mott physics between Sr$_{2}$IrO$_{4}$ and La$_{2}$CuO$_{4}$, 
%leads to the speculation that once doped, 
it was proposed that a spin singlet $d$-wave high temperature (high $T_c$) superconductivity emerges in doped iridates~\cite{Kim12,FaWang11}. 
If this turns out to be true, it would be a significant progress in decades-long efforts to find high $T_c$ superconductivity in other oxides materials besides cuprates. 
On the other hand, doped iridates are inherently multi-orbital systems and the analogy to the cuprates may be justified only in the extremely strong SOC limit.
The determination of the ground states in such multi-orbital systems is a highly challenging theoretical work when the SOC and some of the multi-orbital interactions such as Hund's coupling become comparable to each other, which could easily be the case in 4$d$ or 5$d$ electron systems.

\begin{figure}[h!]
\centering{}
\includegraphics[width=3.3in]{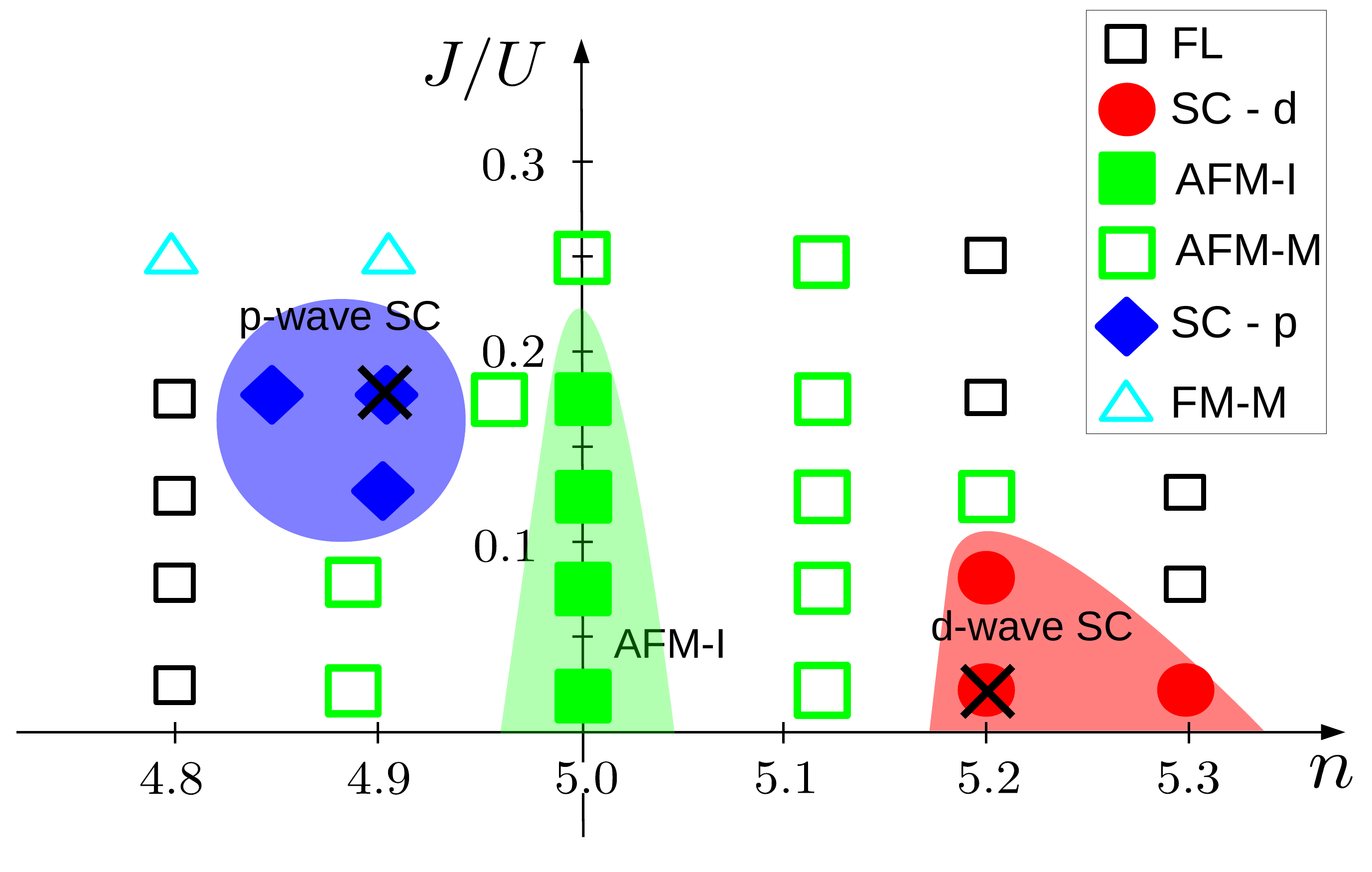}
\caption{(color online) Phase diagram of the $t_{2g}$ Hubbard model in terms of $J/U$ and filling $n$, obtained from DMFT with CTQMC and parquet formulation, where $J$ and $U$ represent the Hund's and intra-orbital Hubbard interaction, respectively. The tight-binding parameters and spin-orbit coupling strength are fixed (see main text), and the lowest temperature achieved in the simulation is $0.05t$. Symbols correspond to the parameter sets where simulations are performed. 
FL, SC-d, AFM-I, AFM-M, SC-p and FM-M stand for Fermi liquid, $d$-wave singlet pairing, antiferromagnetic insulator, antiferromagnetic metal, $p$-wave triplet pairing and ferromagnetic metal, respectively. The shaded areas are guides to the eye, and the two crosses highlight the two selected parameter sets where the instability analyses are presented in Fig.~\ref{fig:electron_dope_lev} and Fig.~\ref{fig:hole_dope_lev}.}
\label{fig:phase_diagram}
\end{figure}

In this letter, we provide a theoretical study of possible superconductivity in $t_{2g}$ multi-orbital systems with SOC using 
the combination of the DMFT with CTQMC impurity solver~\cite{Werner06, Haule07, Huang12a, Du13b} and 
self-consistent relations between two-particle correlation/vertex functions in parquet equations~\cite{Dominicis64a, *Dominicis64b,Bickers91,SXYang2009,Rohringer12,Tam13}. 
The DMFT with CTQMC can capture the local correlation effects, but cannot provide the momentum 
dependence of the vertex functions or two-particle correlation functions, which is necessary for the determination of the dominant
pairing channel and other instabilities. A standard way to introduce the momentum dependence is to generalize the single-site
effective impurity problem to a finite cluster. While the cluster DMFT has been successful for one-band Hubbard models~\cite{Maier05,Gull11,Chen13a,Chen13b},
it would be computationally too costly if one applies it to the multi-orbital models with intra-, inter-orbital interactions and SOC.
Here we use an alternative method via the two-particle diagrammatic relations in the Bethe-Salpeter and parquet equations. 
As described below, we use the results of the DMFT with CTQMC as an input and bring out momentum dependence of necessary 
vertex functions via the relations between vertex and two-particle correlation functions in different interaction channels.

Our major findings are summarized in the phase diagram of Fig.~\ref{fig:phase_diagram}, where
$J$ is the Hund's coupling and $n$ is the band filling. When $J$ becomes comparable to SOC, a two-fold degenerate $p$-wave triplet (in terms of a Kramers-doublet) superconductivity emerges in the hole-doped side, with moderately high transition temperature. On the other hand, $d$-wave superconductivity arises in the electron-doped side when $J$ is small, but is suppressed as $J$ is increased. Note that previous studies reported $d$-wave superconductivity in the electron-doped side~\cite{Arita12,Watanabe13,Wang14} and $s^{*}_{\pm}$-wave in the hole-doped side~\cite{Wang14}, but did not find odd-parity $p$-wave triplet superconductivity. It is also important to emphasize that our odd-parity triplet pairing state is different from the spin-triplet, orbital-singlet pairing state found in previous single-site DMFT studies~\cite{Han04,Sakai04} and mean-field study~\cite{Puetter12}. We show that the emergence of the $p$- and $d$-wave superconducting instabilities in the hole- and electron-doped sides are related to enhanced charge and antiferromagnetic fluctuations, respectively. Below we discuss the microscopic model, numerical method and implications of our results to doped iridates.

\paragraph*{Microscopic Model.-} The $t_{2g}$ three-orbital Hubbard model on the square lattice is given by, $H=H_{kin}+H_{SOC}+H_{I}$, where $H_{kin}=\sum_{\mathbf{k} \alpha \sigma}\epsilon_{\alpha}(\mathbf{k})c^{\dagger}_{\mathbf{k} \alpha \sigma}c_{\mathbf{k} \alpha \sigma}$, $c_{\mathbf{k} \alpha \sigma}$ is the 
electron operator with momentum ${\mathbf{k}}$, spin $\sigma=\uparrow,\downarrow$ and orbital $\alpha=(d_{yz}, d_{zx}, d_{xy})$. 
The SOC term is given by $H_{SOC}=\lambda\sum_{i, \alpha \alpha', \sigma \sigma'} \langle\alpha|\mathbf{L}_{i}|\alpha' 
\rangle\langle\sigma|\mathbf{S}_{i}|\sigma'\rangle c^{\dagger}_{i \alpha \sigma}c_{i \alpha' \sigma'}$, 
and $\mathbf{L}_{i}$($\mathbf{S}_{i}$) is the orbital(spin) angular momentum operator. The interaction term can be written as
\begin{eqnarray}
 H_{I} &=& U\sum_{i, \alpha} n_{i \alpha \uparrow} n_{i \alpha \downarrow} + {U' \over 2} \sum_{i, \alpha \not= \alpha'} n_{i \alpha} n_{i \alpha'} \cr
&+& {J \over 2} \sum_{i, \alpha \not= \alpha', \sigma \sigma'} c^{\dagger}_{i \alpha \sigma} c^{\dagger}_{i \alpha' \sigma'} c_{i \alpha \sigma'} c_{i \alpha' \sigma} \cr
&+& {J' \over 2} \sum_{i, \alpha \not= \alpha'} c^{\dagger}_{i \alpha \uparrow} c^{\dagger}_{i \alpha \downarrow} c_{i \alpha' \downarrow} c_{i \alpha' \uparrow},
\end{eqnarray} 
where  $n_{i \alpha \sigma} = c^{\dagger}_{i\alpha \sigma} c_{i\alpha \sigma}$ and $n_{i \alpha} = \sum_{\sigma} n_{i \alpha \sigma}$. $U'$ and $J'$ denote inter-orbital Hubbard interaction and pair hopping, respectively. In the atomic limit, these four Kanamori parameters satisfy the relation, $U = U' + J + J'$ and $J = J'$, which is assumed in the following discussions. Thus we explore the phase diagram in terms of $U$ and the Hund's coupling $J$.

\begin{figure}[h!]
\centering{}
\includegraphics[width=2.5in]{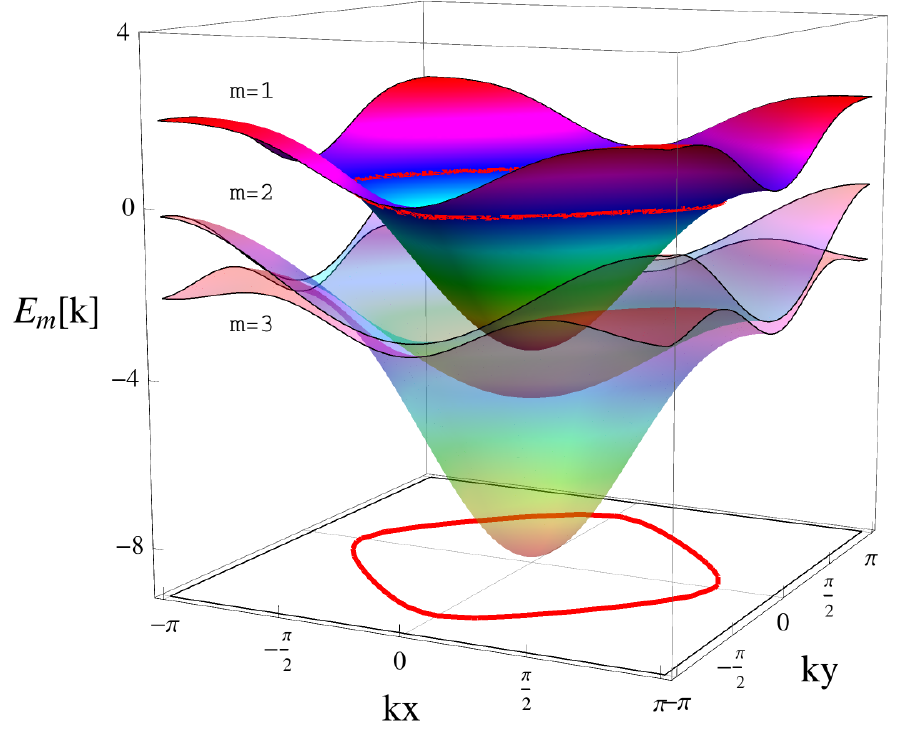}
\caption{(color online) The band dispersion $E_{m}(\mathbf{k})$ and Fermi surface (FS) at filling $n=5$. The spin-orbit coupling $\lambda$ separates the $m=1$ and 
the $m=2,3$ bands. At $\lambda=2t$, the FS only crosses the $m=1$ band, as shown by the red contour line and its projection to the bottom of the BZ.}
\label{fig:dispersion}
\end{figure}

The SOC mixes electron spin and orbital quantum numbers, hence it is useful to first diagonalize the non-interacting Hamiltonian $H_{kin}+H_{SOC}=\sum_{\mathbf{k} m s}E_{m}(\mathbf{k})a^{\dagger}_{\mathbf{k} m s}a_{\mathbf{k} m s}$ (see Supplemental Material~\cite{supplemental} for details), where $a^{\dagger}_{\mathbf{k} m s}$ represent the spin-orbit entangled eigenstates 
characterized by the band index $m=(1,2,3)$ and pseudospin $s$ (a Kramers-doublet) with the dispersion $E_{m}(\mathbf{k})$. 
We adopt the tight-banding parameters of $\epsilon_{\alpha}(\mathbf{k})$ used in Refs.~\cite{Watanabe10,Watanabe13,Wang14,Watanabe14}, the nearest-neighbor hopping between $d_{xy}$ orbitals as the energy unit $t$, and the spin-orbit coupling $\lambda=2t$. 
The energy dispersions $E_{m}(\mathbf{k})$ and the Fermi surface (FS) at filling $n=5$ are shown in Fig.~\ref{fig:dispersion}. The $m=1$ 
band, mostly made of $J_{\text{eff}}=1/2$ state, is separated from the other two 
bands. Near $n=5$ band filling, only $m=1$ band crosses the Fermi level so that there is a single electron-like FS, as shown by the red contour line and its projection to the bottom of the 
Brillouin zone (BZ) in Fig.~\ref{fig:dispersion}. 

\paragraph*{Numerical method.-}
To solve the interacting electron problem,
we employ the DMFT with CTQMC impurity solver~\cite{Maier05,Gull11}. 
This method maps the original, strongly correlated, lattice system into a quantum impurity problem embedded in a self-consistently-determined bath. 
In this study, we use the hybridization expansion CTQMC impurity solver~\cite{Werner06,Haule07,Huang12a}. It diagonalizes the atomic limit of the interacting problem, and diagrammatically expands the impurity partition function in powers of the hybridization function between the impurity and the bath. Since this algorithm treats the local interactions exactly,
it is particularly efficient at moderate and strong interactions. 
We use about $10^{9}$ Monte Carlo samples per simulation to obtain converged single-particle results, and another $10^{9}$ QMC samples to obtain two-particle quantities. The interaction strength is chosen to be close to the bare bandwidth, $U=12t$~\cite{footnote14}, and we can achieve temperatures as low as $T=0.05t$ ($\beta t = 20$) before a serious minus-sign problem renders the data untrustable.

\begin{figure}[ht!]
\centerline{
\includegraphics[width=3.2in]{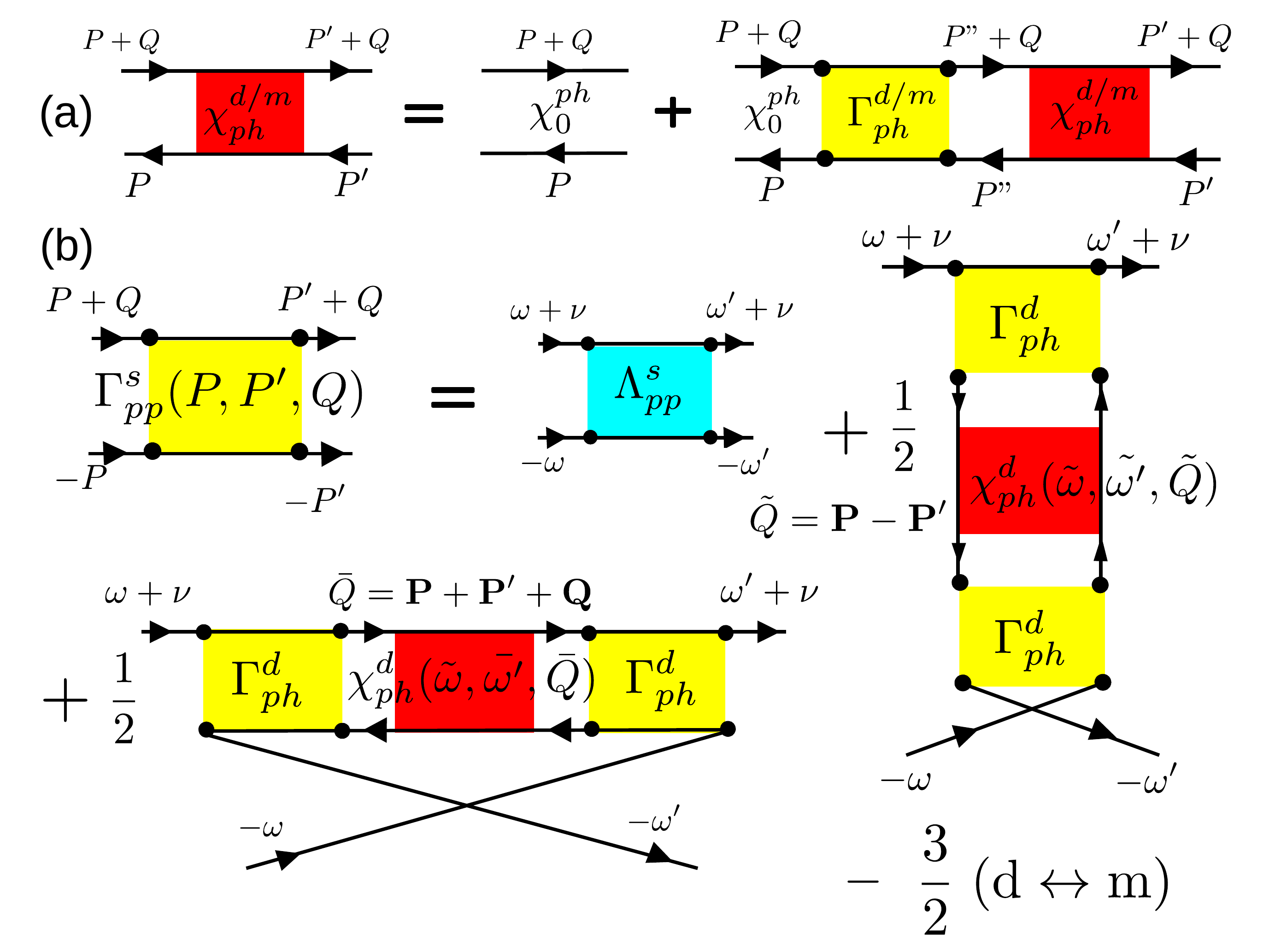}}
\caption{(color online) (a) Bethe-Salpeter equation in the particle-hole density/magnetic channels. $\chi^{d/m}_{ph}$ 
and $\Gamma^{d/m}_{ph}$ are two-particle correlation and vertex functions, and $\chi^{ph}_{0}$ is the bare two-particle
correlation function. 
(b) Parquet equation for the particle-particle singlet vertex, $\Gamma^{s}_{pp}(P,P',Q)$. It is decomposed into fully irreducible vertex function $\Lambda^{s}_{pp}$ and cross-channel contributions from particle-hole density/magnetic vertex ladders $\Phi^{d/m}_{ph}= \Gamma^{d/m}_{ph}\star \chi^{d/m}_{ph} \star \Gamma^{d/m}_{ph}$ (complete equations are given in Supplemental Material~\cite{supplemental}).}
\label{fig:parquet_pp}
\end{figure}

In order to obtain information about pairing instabilities, one needs to know the momentum and frequency dependence of the pairing vertex functions.
However, in the DMFT simulation, the two-particle correlation functions in the particle-particle (pp) and particle-hole (ph) channels $\chi_{ph/pp}(\omega,\omega',\nu)$ can only be measured at the impurity site, hence only have frequency-dependence. Here we use the parquet equations to introduce momentum-dependence in two-particle quantities as described below. The parquet equations relate the irreducible vertex function in one interaction channel to those in other channels~\cite{Dominicis64a, *Dominicis64b}. 
In our case, we consider four interaction channels: the particle-hole density (ph-d), particle-hole magnetic (ph-m), particle-particle singlet (pp-s) and particle-particle triplet (pp-t) channels~\cite{Bickers91,SXYang2009,Rohringer12,Tam13,Chen13a,Chen13b}. A detailed description of the parquet formalism is given in the Supplemental Material~\cite{supplemental} and here we only outline the main idea. 

For example, in order to explore the singlet/triplet paring instabilities, we need to find the momentum and frequency dependence of the irreducible vertex functions in pp-s/t channels,
$\Gamma^{s/t}_{pp} (P,P',Q)$, with $P\equiv(\mathbf{k},\omega)$, $P'\equiv(\mathbf{k'},\omega')$, $Q\equiv(\mathbf{q},\nu)$.
In the DMFT-CTQMC, one obtains the lattice single-particle Green's function $G(P)$ and the ph-d/m two-particle correlation functions $\chi^{d/m}_{ph}(\omega,\omega',\nu)$ measured on the impurity. We first consider the local version of the Bethe-Salpeter equation, $\chi^{d/m}_{ph}(\omega,\omega',\nu)=\chi^{ph}_{0}(\omega,\nu)+\chi^{ph}_{0}(\omega,\nu)\sum_{\omega''}\Gamma^{d/m}_{ph}(\omega,\omega'',\nu)\chi^{d/m}_{ph}(\omega'',\omega',\nu)$.
Using $\chi^{d/m}_{ph}(\omega,\omega',\nu)$ obtained in the DMFT, one can extract the local irreducible vertex functions, $\Gamma^{d/m}_{ph}(\omega,\omega'',\nu)$. 

To introduce the momentum dependence in the vertex functions starting from $G(P)$, $\chi^{d/m}_{ph}(\omega,\omega',\nu)$, and
$\Gamma^{d/m}_{ph}(\omega,\omega'',\nu)$, let us turn to the lattice Bethe-Salpeter equation in Fig.~\ref{fig:parquet_pp} (a); 
$\chi^{d/m}_{ph}(P,P',Q) = \chi^{ph}_{0}(P,Q)+
\chi^{ph}_{0}(P,Q)\sum_{P''} \Gamma^{d/m}_{ph} (P,P'',Q)\chi^{d/m}_{ph}(P'',P',Q)$,
where $\chi^{ph}_{0}(P,Q)$ can be constructed from single-particle Green's functions 
$\chi^{ph}_{0}(P,Q)=-N\beta G(P)G(P+Q)$ with $N$, the lattice size. We then use $\Gamma^{d/m}_{ph} (\omega,\omega'',\nu)$ (obtained in the DMFT) as an input for $\Gamma^{d/m}_{ph} (P,P'',Q)$,
and later find the momentum dependence of this and other quantities using an iteration method.
Once $\Gamma^{d/m}_{ph} (\omega,\omega'',\nu)$ is used and the sums over ${\bf k}$, ${\bf k'}$ are applied to both
sides of the equation, the Bethe-Salpeter equation is reduced to
\begin{eqnarray}
\chi^{d/m}_{ph}(\omega,\omega',Q)&=&\chi^{ph}_{0}(\omega,Q)+\nonumber\\
\chi^{ph}_{0}(\omega,Q)\sum_{\omega''}&\Gamma^{d/m}_{ph}&(\omega,\omega'',\nu)\chi^{d/m}_{ph}(\omega'',\omega',Q),
\label{eq:Bethe_Salpeter_2}
\end{eqnarray}
where 
$\chi^{ph}_{0}(\omega,Q) = \sum_{\bf k} \chi^{ph}_0 (P,Q)$
and $\chi^{d/m}_{ph}(\omega,\omega',Q)=\sum_{{\bf k},{\bf k}'} \chi^{d/m}_{ph}(P,P',Q)$. $\chi^{d/m}_{ph}(\omega,\omega',Q)$ is then obtained by solving Eq.~\ref{eq:Bethe_Salpeter_2}.

Now we consider the parquet equation in Fig.~\ref{fig:parquet_pp} (b), where the irreducible vertex functions in the pp channel, $\Gamma^{s/t}_{pp} (P,P',Q)$, are
related to $\Gamma^{d/m}_{ph} (P,P',Q)$ and $\chi^{d/m}_{ph} (P,P',Q)$ via the ph vertex ladders $\Phi^{d/m}_{ph} = 
\Gamma^{d/m}_{ph} \star \chi^{d/m}_{ph} \star \Gamma^{d/m}_{ph}$, where $\star$ represents the convolution in
both momentum and frequency. In order to get the first order results for $\Gamma^{(1), s/t}_{pp} (P,P',Q)$, we use 
$\Gamma^{d/m}_{ph} (\omega,\omega',\nu)$ and $\chi^{d/m}_{ph}(\omega,\omega',{\tilde Q})$ for
$\Gamma^{d/m}_{ph} (P,P',Q)$ and $\chi^{d/m}_{ph}(P,P',{\tilde Q})$ in the ph ladders $\Phi^{d/m}_{ph}$ with
the momentum-frequency convolution replaced by a frequency-only convolution.
Here ${\tilde Q} = P-P'$ or $P+P'+Q$, which provides the momentum dependence in $\Gamma^{(1), s/t}_{pp} (P,P',Q)$.
Similar procedure is employed to get $\Gamma^{(1), d/m}_{ph} (P,P',Q)$.
These first order results are now iterated back to the full Bethe-Salpeter and parquet equations, and then successive iterations would generate higher order results.
In principle, this procedure needs to be repeated until self-consistency is achieved. Such calculations, however, require 
unrealistic amount of computing resources. Instead, we check explicitly that the results of $\Gamma^{(1)}$, $\Gamma^{(2)}$ and $\Gamma^{(3)}$ are consistent with each other and, as shown later, provide the same trend in the instability analysis for various interaction channels (in fact, the results are almost converged at $\Gamma^{(3)}$, see Supplemental Material~\cite{supplemental}).

%various interaction channels .in we use the results of $\Gamma^{(2)}$ to 
%investigate instabilities in various interaction channels.

\begin{figure}[tp!]
\centerline{
\includegraphics[width=3.3in]{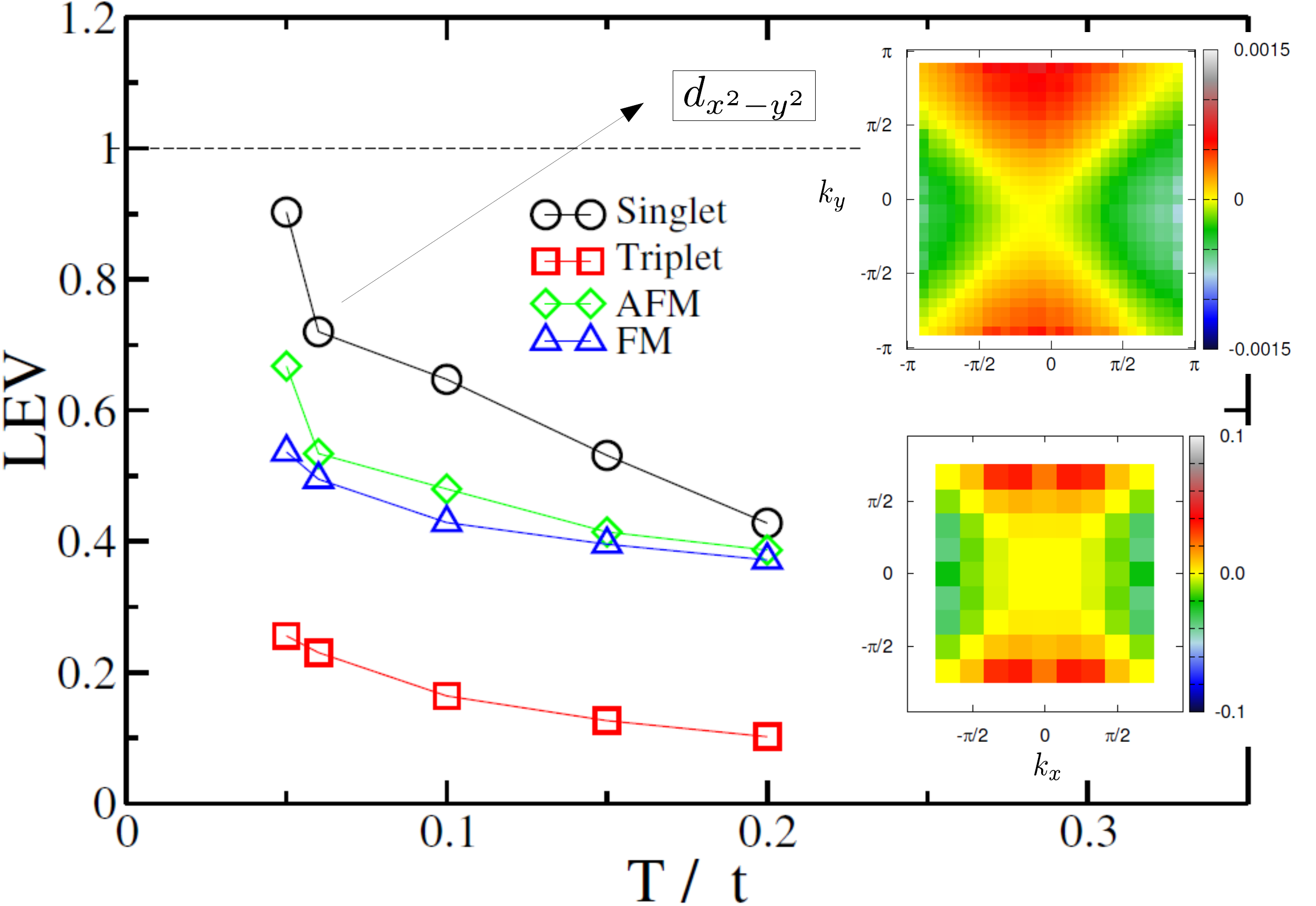}}
  \caption{(color online) The leading eigenvalues (LEV) of $\Gamma^{(1)}$ as a function of temperature in various instability channels, for 
  an electron-doped case ($n=5.2$) with a small Hund's coupling ($J=0.2t$) and $U=12t$, $U'=11.6t$.
  The upper(lower) inset shows the $d$-wave symmetry of the leading eigenvector in the singlet pairing channel for $\Gamma^{(1)}$($\Gamma^{(2)}$).}
  \label{fig:electron_dope_lev}
\end{figure}

For instance, we use the irreducible vertex functions $\Gamma^{(1),s/t}_{pp}(P,P',Q)$ and/or $\Gamma^{(2),s/t}_{pp}(P,P',Q)$ to study superconducting instabilities via
\begin{equation}
\sum_{P'}\Gamma^{s/t}_{pp}(P,P',Q)\chi^{pp}_0(P',Q)\phi(P') = \lambda \phi(P),
\label{eq:pairing_eigen_equation}
\end{equation}
where the leading eigenvalue (LEV) $\lambda$ and the leading eigenvector $\phi(P)$ need to be analyzed. 
As temperature approaches the transition temperature $T_{c}$, $\lambda \to 1$, and the corresponding $\phi(P)$ shows 
the momentum-dependence of the gap function~\cite{Schrieffer83,Chen13a}. Similar analysis can be performed in the ph-d/m channels.

\paragraph*{Results and Discussions.-}
We compute the LEVs of $\Gamma^{(1)}$ and $\Gamma^{(2)}$ for the corresponding vertex functions as a function of temperature $T$ for singlet/triplet superconductivity, ferromagnetic and antiferromagnetic instabilities across the phase diagram and the leading eigenvector is used to determine the ground state.
Fig.~\ref{fig:electron_dope_lev} shows the results for the parameter set $U=12t$, $U'=11.6t$, $J=0.2t$, $n=5.2$. This is an electron-doped case with a very small Hund's coupling $J/U \sim 0.017$. The main panel shows the LEVs obtained from Eq.~\ref{eq:pairing_eigen_equation} using $\Gamma^{(1)}$ (the results of $\Gamma^{(2)}$ show the same trend). As temperature decreases, the (pseudospin-)singlet pairing LEV in the $m=1$ band dominates over other channels and the antiferromagnetic channel is the next leading instability. Moreover, the leading eigenvector of the singlet pairing clearly has the $d_{x^2-y^2}$ momentum-dependence, as shown in the upper inset (the lower inset shows the leading eigenvector of $\Gamma^{(2)}$ which has the same $d$-wave symmetry). In the electron-doped side, both Hund's coupling and SOC prefer to have $J_{\text{eff}}=3/2$ bands completely filled, and extra electron goes to the initially half-filled $J_{\text{eff}}=1/2$ band. Thus the d-wave singlet pairing mainly comes from the $J_{\text{eff}}=1/2$ band. Moreover, the corresponding FS is very similar to that of the hole-doped, one-band Hubbard model on square lattice.
%Such results are consistent with previous studies of a similar model~\cite{Watanabe13,Wang14} using different methods. 
%Notice that the FS of the three-orbital model in the electron-doped side is topologically similar to the FS of the hoped-dope case in the one-band Hubbard model. 
As shown in the cluster DMFT computations of one-band Hubbard model, the vertex function for the 
$d$-wave superconducting instability is dominated by antiferromagnetic fluctuations~\cite{Chen13b}. Our analysis of the parquet equation shows
that the magnetic vertex ladder $\Phi^{m}_{ph}$ at $\mathbf{q}=(\pi,\pi)$ is indeed the dominant contribution to $\Gamma^{s}_{pp}$.

\begin{figure}[tp!]
\centerline{
  \includegraphics[width=3.3in]{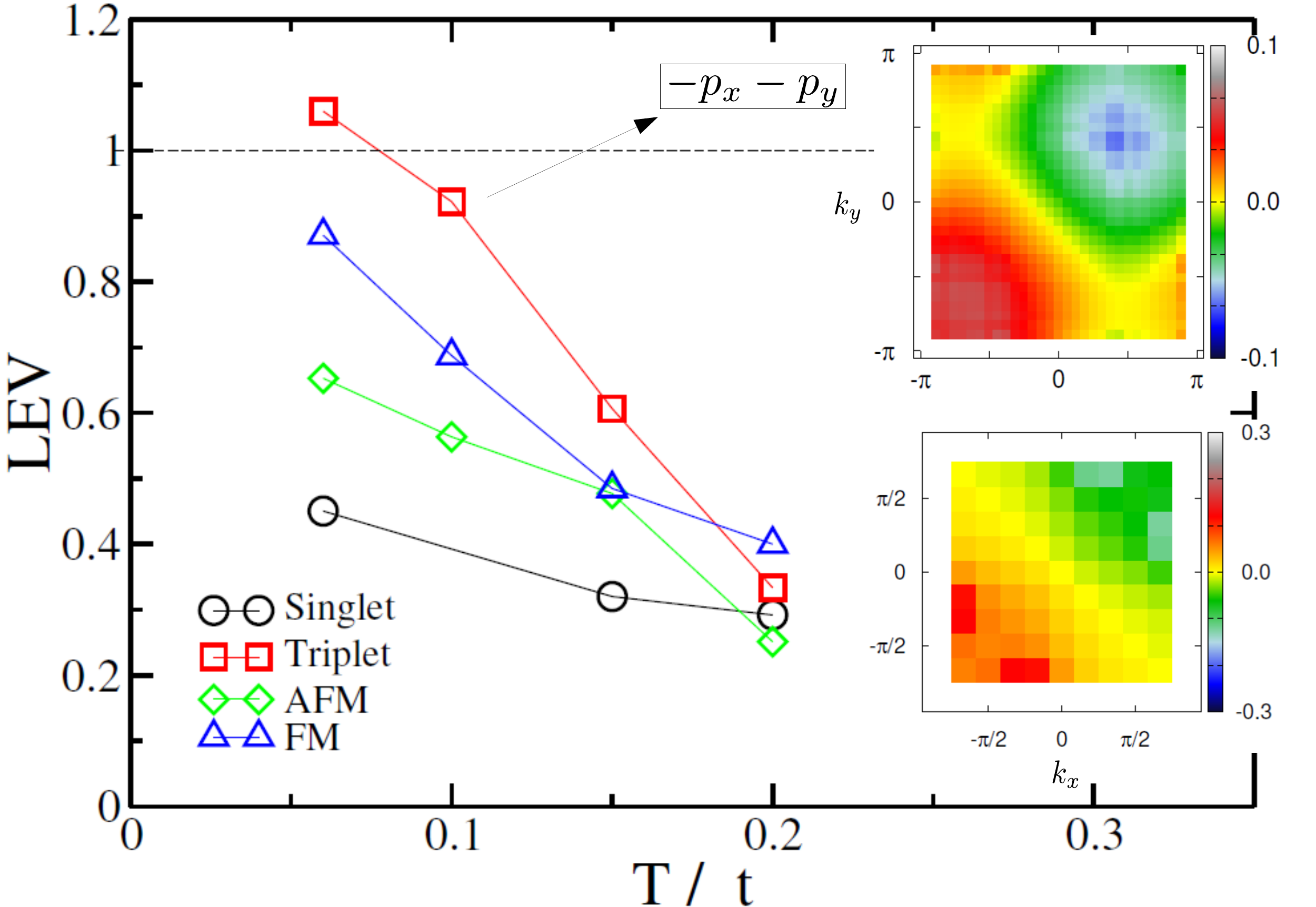}}
  \caption{(color online) 
  The leading eigenvalues (LEV) of $\Gamma^{(1)}$ as a function of temperature in various instability channels, for 
  a hole-doped case ($n=4.9$) with a large Hund's coupling ($J=2t$) and $U=12t$, $U'=8t$.
  The upper(lower) inset shows the $p'_{x}=-p_{x}-p_{y}$ symmetry of the leading eigenvector in the triplet channel for $\Gamma^{(1)}$($\Gamma^{(2)}$).
  The other degenerate $p$-wave component, $p'_{y}=-p_{x}+p_{y}$, is not shown.}
  \label{fig:hole_dope_lev}
\end{figure} 

In turn, the main panel of Fig.~\ref{fig:hole_dope_lev} shows the LEVs of $\Gamma^{(1)}$ at a large Hund's coupling ($J/U \sim 0.17$) in a hole-doped case, with the parameter set $U=12t$, $U'=8t$, $J=2t$, $n=4.9$. As the Hund's coupling increases, the (pseudospin-)triplet pairing in the $m=1$ band becomes the leading instability in the hole-doped side while the $d$-wave singlet pairing in the electron-doped side is suppressed. The triplet pairing instability found here has two-fold degenerate LEVs and the corresponding leading eigenvectors have $p'_{x}=-p_{x}-p_{y}$ and $p'_{y}=-p_{x}+p_{y}$ symmetries. The upper(lower) inset of Fig.~\ref{fig:hole_dope_lev} shows the leading eigenvector obtained from $\Gamma^{(1)}$($\Gamma^{(2)}$) with $p'_{x}$ symmetry. These results imply that the triplet superconductivity is the dominant instability in the hole-doped side. This $p$-wave triplet superconductivity emerges from a delicate balance between SOC and Hund's coupling~\cite{Puetter12}. When holes are introduced, the Hund's coupling prefers to have holes in $J_{\text{eff}}=3/2$ bands as well as $J_{\text{eff}}=1/2$ band, while the spin-orbit coupling likes to have $J_{\text{eff}}=3/2$ completely filled and to put all extra holes in the $J_{\text{eff}}=1/2$ band. Thus two interactions are not compatible to each other. Only when the SOC and Hund's coupling are balanced, ferromagnetic fluctuation induced by Hund's coupling generates the triplet pairing state. If the Hund's coupling becomes even larger, as shown in the phase diagram (Fig.1), in the hole-doped side, the system becomes a ferromagnetic metal. Thus we need a significant Hund's coupling to induce the triplet pairing via ferromagnetic fluctuations, but not-too-large Hund's coupling which eventually favors a ferromagnetic metal.

The odd-parity triplet pairing is doubly degenerate with components $p'_{x}$ and $p'_{y}$, any linear combination of both $p$-wave components is possible below $T_c$. Given that both Ginzburg-Landau theory and BCS-type mean-field approaches favor a fully-gapped superconducting phase that breaks time-reversal symmetry~\cite{Sigrist91},  the $p_x+ip_y$ triplet pairing state could be selected.
%suggested for Sr$_{2}$RuO$_{4}$~\cite{Mackenzie03}, or in 2D lattice systems with type-II van Hove singularities~\cite{Yao13,Ma14}. 
Therefore, our findings may support the chiral $p_x+ip_y$ topological superconducting phase in the hole-doped side of the phase diagram. 
%At larger Hund's coupling, the ferromagnetic instability becomes the leading instability, indicating that subtle optimization between SOC and the Hund's coupling leads to spin-triplet superconductivity~\cite{Puetter12}. The corresponding vertex ladders for the pairing and magnetic channels show that both the triplet pairing and ferromagnetic instabilities are driven by enhanced charge fluctuations with the momentum $\mathbf{q}=(0,0)$. 

It is clear from Fig.~\ref{fig:electron_dope_lev} and Fig.~\ref{fig:hole_dope_lev} that the triplet pairing transition temperature in the hole-doped side is higher than the singlet pairing one in the electron-doped side. That is, the triplet LEVs approaches 1 when  $0.06t \lesssim T \lesssim 0.1t$ whereas the singlet LEV is still below 1 at $T=0.05t$. The same behaviors also hold for the LEVs obtained from $\Gamma^{(2)}$ analyses. This implies that the $p$-wave superconductivity in the hope-doped side could have relatively higher $T_c$ than the $d$-wave superconductivity in the electron-doped side. Although superconductivity has not been observed in electron-doped Sr$_{2}$IrO$_{4}$~\cite{Korneta10}, our results could stimulate more experimental 
efforts in the hole-doped side, which may be achieved by substituting Na, K for Sr.

%\paragraph*{Conclusion.-}
%Using large-scale CTQMC+DMFT+Parquet simulations, we explore the SC instabilities and pairing mechanism in $t_{2g}$ three-orbital Hubbard model. At small Hund's coupling, with electron doping, we find a $d$-wave singlet pairing state among the $J_{\text{eff}}=1/2$ bands, triggered by antiferromagnetic fluctuations. At large Hund's coupling, with hole doping, we find a doubly degenerate $p$-wave triplet pairing state among the $J_{\text{eff}}=1/2$ bands, with higher transition temperature than the $d$-wave singlet pairing, and is associated with the enhanced charge fluctuations. Our findings suggest a possibile, high-$T_{c}$, chiral, topological $p+ip$ triplet superconducting phase in this model. Furthermore, the numerical formalism developed in this work opens promising avenue for future study of correlated multi-orbital systems with spin-orbital coupling.

\paragraph*{Acknowledgements.-}
We would like to acknowledge H. Li, Y. L. Wang for discussions on the DMFT+CTQMC simulation code -- the iQIST package~\cite{iQIST14}, K. S. Chen, S. X. Yang for their help on the parquet formalism, and Q.-H. Wang for discussions on possible pairing symmetries and their fRG work~\cite{Wang14}. This work is supported by the NSERC, CIFAR, and Centre for Quantum Materials at the University of Toronto. Computations were performed on the GPC supercomputer at the SciNet HPC Consortium. SciNet is funded by: the Canada Foundation for Innovation under the auspices of Compute Canada; the Government of Ontario; Ontario Research Fund - Research Excellence; and the University of Toronto.

\bibliography{t2g_Sr2IrO4_Superconductivity_0922}

\begin{thebibliography}{39}%
\makeatletter
\providecommand \@ifxundefined [1]{%
 \@ifx{#1\undefined}
}%
\providecommand \@ifnum [1]{%
 \ifnum #1\expandafter \@firstoftwo
 \else \expandafter \@secondoftwo
 \fi
}%
\providecommand \@ifx [1]{%
 \ifx #1\expandafter \@firstoftwo
 \else \expandafter \@secondoftwo
 \fi
}%
\providecommand \natexlab [1]{#1}%
\providecommand \enquote  [1]{``#1''}%
\providecommand \bibnamefont  [1]{#1}%
\providecommand \bibfnamefont [1]{#1}%
\providecommand \citenamefont [1]{#1}%
\providecommand \href@noop [0]{\@secondoftwo}%
\providecommand \href [0]{\begingroup \@sanitize@url \@href}%
\providecommand \@href[1]{\@@startlink{#1}\@@href}%
\providecommand \@@href[1]{\endgroup#1\@@endlink}%
\providecommand \@sanitize@url [0]{\catcode `\\12\catcode `\$12\catcode
  `\&12\catcode `\#12\catcode `\^12\catcode `\_12\catcode `\%12\relax}%
\providecommand \@@startlink[1]{}%
\providecommand \@@endlink[0]{}%
\providecommand \url  [0]{\begingroup\@sanitize@url \@url }%
\providecommand \@url [1]{\endgroup\@href {#1}{\urlprefix }}%
\providecommand \urlprefix  [0]{URL }%
\providecommand \Eprint [0]{\href }%
\providecommand \doibase [0]{http://dx.doi.org/}%
\providecommand \selectlanguage [0]{\@gobble}%
\providecommand \bibinfo  [0]{\@secondoftwo}%
\providecommand \bibfield  [0]{\@secondoftwo}%
\providecommand \translation [1]{[#1]}%
\providecommand \BibitemOpen [0]{}%
\providecommand \bibitemStop [0]{}%
\providecommand \bibitemNoStop [0]{.\EOS\space}%
\providecommand \EOS [0]{\spacefactor3000\relax}%
\providecommand \BibitemShut  [1]{\csname bibitem#1\endcsname}%
\let\auto@bib@innerbib\@empty
%</preamble>
\bibitem [{\citenamefont {Witczak-Krempa}\ \emph {et~al.}(2014)\citenamefont
  {Witczak-Krempa}, \citenamefont {Chen}, \citenamefont {Kim},\ and\
  \citenamefont {Balents}}]{Krempa14}%
  \BibitemOpen
  \bibfield  {author} {\bibinfo {author} {\bibfnamefont {W.}~\bibnamefont
  {Witczak-Krempa}}, \bibinfo {author} {\bibfnamefont {G.}~\bibnamefont
  {Chen}}, \bibinfo {author} {\bibfnamefont {Y.~B.}\ \bibnamefont {Kim}}, \
  and\ \bibinfo {author} {\bibfnamefont {L.}~\bibnamefont {Balents}},\ }\href
  {\doibase 10.1146/annurev-conmatphys-020911-125138} {\bibfield  {journal}
  {\bibinfo  {journal} {Annu. Rev. Condens. Matter Phys.}\ }\textbf {\bibinfo
  {volume} {5}},\ \bibinfo {pages} {57} (\bibinfo {year} {2014})}\BibitemShut
  {NoStop}%
\bibitem [{\citenamefont {Cao}\ \emph {et~al.}(1998)\citenamefont {Cao},
  \citenamefont {Bolivar}, \citenamefont {McCall}, \citenamefont {Crow},\ and\
  \citenamefont {Guertin}}]{Cao98}%
  \BibitemOpen
  \bibfield  {author} {\bibinfo {author} {\bibfnamefont {G.}~\bibnamefont
  {Cao}}, \bibinfo {author} {\bibfnamefont {J.}~\bibnamefont {Bolivar}},
  \bibinfo {author} {\bibfnamefont {S.}~\bibnamefont {McCall}}, \bibinfo
  {author} {\bibfnamefont {J.~E.}\ \bibnamefont {Crow}}, \ and\ \bibinfo
  {author} {\bibfnamefont {R.~P.}\ \bibnamefont {Guertin}},\ }\href {\doibase
  10.1103/PhysRevB.57.R11039} {\bibfield  {journal} {\bibinfo  {journal} {Phys.
  Rev. B}\ }\textbf {\bibinfo {volume} {57}},\ \bibinfo {pages} {R11039}
  (\bibinfo {year} {1998})}\BibitemShut {NoStop}%
\bibitem [{\citenamefont {Kim}\ \emph {et~al.}(2008)\citenamefont {Kim},
  \citenamefont {Jin}, \citenamefont {Moon}, \citenamefont {Kim}, \citenamefont
  {Park}, \citenamefont {Leem}, \citenamefont {Yu}, \citenamefont {Noh},
  \citenamefont {Kim}, \citenamefont {Oh}, \citenamefont {Park}, \citenamefont
  {Durairaj}, \citenamefont {Cao},\ and\ \citenamefont {Rotenberg}}]{BJKim08}%
  \BibitemOpen
  \bibfield  {author} {\bibinfo {author} {\bibfnamefont {B.~J.}\ \bibnamefont
  {Kim}}, \bibinfo {author} {\bibfnamefont {H.}~\bibnamefont {Jin}}, \bibinfo
  {author} {\bibfnamefont {S.~J.}\ \bibnamefont {Moon}}, \bibinfo {author}
  {\bibfnamefont {J.-Y.}\ \bibnamefont {Kim}}, \bibinfo {author} {\bibfnamefont
  {B.-G.}\ \bibnamefont {Park}}, \bibinfo {author} {\bibfnamefont {C.~S.}\
  \bibnamefont {Leem}}, \bibinfo {author} {\bibfnamefont {J.}~\bibnamefont
  {Yu}}, \bibinfo {author} {\bibfnamefont {T.~W.}\ \bibnamefont {Noh}},
  \bibinfo {author} {\bibfnamefont {C.}~\bibnamefont {Kim}}, \bibinfo {author}
  {\bibfnamefont {S.-J.}\ \bibnamefont {Oh}}, \bibinfo {author} {\bibfnamefont
  {J.-H.}\ \bibnamefont {Park}}, \bibinfo {author} {\bibfnamefont
  {V.}~\bibnamefont {Durairaj}}, \bibinfo {author} {\bibfnamefont
  {G.}~\bibnamefont {Cao}}, \ and\ \bibinfo {author} {\bibfnamefont
  {E.}~\bibnamefont {Rotenberg}},\ }\href {\doibase
  10.1103/PhysRevLett.101.076402} {\bibfield  {journal} {\bibinfo  {journal}
  {Phys. Rev. Lett.}\ }\textbf {\bibinfo {volume} {101}},\ \bibinfo {pages}
  {076402} (\bibinfo {year} {2008})}\BibitemShut {NoStop}%
\bibitem [{\citenamefont {Kim}\ \emph {et~al.}(2009)\citenamefont {Kim},
  \citenamefont {Ohsumi}, \citenamefont {Komesu}, \citenamefont {Sakai},
  \citenamefont {Morita}, \citenamefont {Takagi},\ and\ \citenamefont
  {Arima}}]{BJKim09}%
  \BibitemOpen
  \bibfield  {author} {\bibinfo {author} {\bibfnamefont {B.~J.}\ \bibnamefont
  {Kim}}, \bibinfo {author} {\bibfnamefont {H.}~\bibnamefont {Ohsumi}},
  \bibinfo {author} {\bibfnamefont {T.}~\bibnamefont {Komesu}}, \bibinfo
  {author} {\bibfnamefont {S.}~\bibnamefont {Sakai}}, \bibinfo {author}
  {\bibfnamefont {T.}~\bibnamefont {Morita}}, \bibinfo {author} {\bibfnamefont
  {H.}~\bibnamefont {Takagi}}, \ and\ \bibinfo {author} {\bibfnamefont
  {T.}~\bibnamefont {Arima}},\ }\href@noop {} {\bibfield  {journal} {\bibinfo
  {journal} {Science}\ }\textbf {\bibinfo {volume} {323}},\ \bibinfo {pages}
  {1329} (\bibinfo {year} {2009})}\BibitemShut {NoStop}%
\bibitem [{\citenamefont {Fujiyama}\ \emph {et~al.}(2012)\citenamefont
  {Fujiyama}, \citenamefont {Ohsumi}, \citenamefont {Komesu}, \citenamefont
  {Matsuno}, \citenamefont {Kim}, \citenamefont {Takata}, \citenamefont
  {Arima},\ and\ \citenamefont {Takagi}}]{Fujiyama12}%
  \BibitemOpen
  \bibfield  {author} {\bibinfo {author} {\bibfnamefont {S.}~\bibnamefont
  {Fujiyama}}, \bibinfo {author} {\bibfnamefont {H.}~\bibnamefont {Ohsumi}},
  \bibinfo {author} {\bibfnamefont {T.}~\bibnamefont {Komesu}}, \bibinfo
  {author} {\bibfnamefont {J.}~\bibnamefont {Matsuno}}, \bibinfo {author}
  {\bibfnamefont {B.~J.}\ \bibnamefont {Kim}}, \bibinfo {author} {\bibfnamefont
  {M.}~\bibnamefont {Takata}}, \bibinfo {author} {\bibfnamefont
  {T.}~\bibnamefont {Arima}}, \ and\ \bibinfo {author} {\bibfnamefont
  {H.}~\bibnamefont {Takagi}},\ }\href {\doibase
  10.1103/PhysRevLett.108.247212} {\bibfield  {journal} {\bibinfo  {journal}
  {Phys. Rev. Lett.}\ }\textbf {\bibinfo {volume} {108}},\ \bibinfo {pages}
  {247212} (\bibinfo {year} {2012})}\BibitemShut {NoStop}%
\bibitem [{\citenamefont {Kim}\ \emph {et~al.}(2012)\citenamefont {Kim},
  \citenamefont {Casa}, \citenamefont {Upton}, \citenamefont {Gog},
  \citenamefont {Kim}, \citenamefont {Mitchell}, \citenamefont {van
  Veenendaal}, \citenamefont {Daghofer}, \citenamefont {van~den Brink},
  \citenamefont {Khaliullin},\ and\ \citenamefont {Kim}}]{Kim12}%
  \BibitemOpen
  \bibfield  {author} {\bibinfo {author} {\bibfnamefont {J.}~\bibnamefont
  {Kim}}, \bibinfo {author} {\bibfnamefont {D.}~\bibnamefont {Casa}}, \bibinfo
  {author} {\bibfnamefont {M.~H.}\ \bibnamefont {Upton}}, \bibinfo {author}
  {\bibfnamefont {T.}~\bibnamefont {Gog}}, \bibinfo {author} {\bibfnamefont
  {Y.-J.}\ \bibnamefont {Kim}}, \bibinfo {author} {\bibfnamefont {J.~F.}\
  \bibnamefont {Mitchell}}, \bibinfo {author} {\bibfnamefont {M.}~\bibnamefont
  {van Veenendaal}}, \bibinfo {author} {\bibfnamefont {M.}~\bibnamefont
  {Daghofer}}, \bibinfo {author} {\bibfnamefont {J.}~\bibnamefont {van~den
  Brink}}, \bibinfo {author} {\bibfnamefont {G.}~\bibnamefont {Khaliullin}}, \
  and\ \bibinfo {author} {\bibfnamefont {B.~J.}\ \bibnamefont {Kim}},\ }\href
  {\doibase 10.1103/PhysRevLett.108.177003} {\bibfield  {journal} {\bibinfo
  {journal} {Phys. Rev. Lett.}\ }\textbf {\bibinfo {volume} {108}},\ \bibinfo
  {pages} {177003} (\bibinfo {year} {2012})}\BibitemShut {NoStop}%
\bibitem [{\citenamefont {Carter}\ \emph {et~al.}(2012)\citenamefont {Carter},
  \citenamefont {Shankar}, \citenamefont {Zeb},\ and\ \citenamefont
  {Kee}}]{Carter12}%
  \BibitemOpen
  \bibfield  {author} {\bibinfo {author} {\bibfnamefont {J.-M.}\ \bibnamefont
  {Carter}}, \bibinfo {author} {\bibfnamefont {V.~V.}\ \bibnamefont {Shankar}},
  \bibinfo {author} {\bibfnamefont {M.~A.}\ \bibnamefont {Zeb}}, \ and\
  \bibinfo {author} {\bibfnamefont {H.-Y.}\ \bibnamefont {Kee}},\ }\href
  {\doibase 10.1103/PhysRevB.85.115105} {\bibfield  {journal} {\bibinfo
  {journal} {Phys. Rev. B}\ }\textbf {\bibinfo {volume} {85}},\ \bibinfo
  {pages} {115105} (\bibinfo {year} {2012})}\BibitemShut {NoStop}%
\bibitem [{\citenamefont {Carter}\ \emph {et~al.}(2013)\citenamefont {Carter},
  \citenamefont {Shankar~V.},\ and\ \citenamefont {Kee}}]{Carter13}%
  \BibitemOpen
  \bibfield  {author} {\bibinfo {author} {\bibfnamefont {J.-M.}\ \bibnamefont
  {Carter}}, \bibinfo {author} {\bibfnamefont {V.}~\bibnamefont {Shankar~V.}},
  \ and\ \bibinfo {author} {\bibfnamefont {H.-Y.}\ \bibnamefont {Kee}},\ }\href
  {\doibase 10.1103/PhysRevB.88.035111} {\bibfield  {journal} {\bibinfo
  {journal} {Phys. Rev. B}\ }\textbf {\bibinfo {volume} {88}},\ \bibinfo
  {pages} {035111} (\bibinfo {year} {2013})}\BibitemShut {NoStop}%
\bibitem [{\citenamefont {Ye}\ \emph {et~al.}(2013)\citenamefont {Ye},
  \citenamefont {Chi}, \citenamefont {Chakoumakos}, \citenamefont
  {Fernandez-Baca}, \citenamefont {Qi},\ and\ \citenamefont {Cao}}]{Ye13}%
  \BibitemOpen
  \bibfield  {author} {\bibinfo {author} {\bibfnamefont {F.}~\bibnamefont
  {Ye}}, \bibinfo {author} {\bibfnamefont {S.}~\bibnamefont {Chi}}, \bibinfo
  {author} {\bibfnamefont {B.~C.}\ \bibnamefont {Chakoumakos}}, \bibinfo
  {author} {\bibfnamefont {J.~A.}\ \bibnamefont {Fernandez-Baca}}, \bibinfo
  {author} {\bibfnamefont {T.}~\bibnamefont {Qi}}, \ and\ \bibinfo {author}
  {\bibfnamefont {G.}~\bibnamefont {Cao}},\ }\href {\doibase
  10.1103/PhysRevB.87.140406} {\bibfield  {journal} {\bibinfo  {journal} {Phys.
  Rev. B}\ }\textbf {\bibinfo {volume} {87}},\ \bibinfo {pages} {140406}
  (\bibinfo {year} {2013})}\BibitemShut {NoStop}%
\bibitem [{\citenamefont {Li}\ \emph {et~al.}(2013)\citenamefont {Li},
  \citenamefont {Cao}, \citenamefont {Okamoto}, \citenamefont {Yi},
  \citenamefont {Lin}, \citenamefont {Sales}, \citenamefont {Yan},
  \citenamefont {Arita}, \citenamefont {Kunes}, \citenamefont {Kozhevnikov},
  \citenamefont {Eguiluz}, \citenamefont {Imada}, \citenamefont {Gai},
  \citenamefont {Pan},\ and\ \citenamefont {Mandrus}}]{Li13}%
  \BibitemOpen
  \bibfield  {author} {\bibinfo {author} {\bibfnamefont {Q.}~\bibnamefont
  {Li}}, \bibinfo {author} {\bibfnamefont {G.}~\bibnamefont {Cao}}, \bibinfo
  {author} {\bibfnamefont {S.}~\bibnamefont {Okamoto}}, \bibinfo {author}
  {\bibfnamefont {J.}~\bibnamefont {Yi}}, \bibinfo {author} {\bibfnamefont
  {W.}~\bibnamefont {Lin}}, \bibinfo {author} {\bibfnamefont {B.~C.}\
  \bibnamefont {Sales}}, \bibinfo {author} {\bibfnamefont {J.}~\bibnamefont
  {Yan}}, \bibinfo {author} {\bibfnamefont {R.}~\bibnamefont {Arita}}, \bibinfo
  {author} {\bibfnamefont {J.}~\bibnamefont {Kunes}}, \bibinfo {author}
  {\bibfnamefont {A.~V.}\ \bibnamefont {Kozhevnikov}}, \bibinfo {author}
  {\bibfnamefont {A.~G.}\ \bibnamefont {Eguiluz}}, \bibinfo {author}
  {\bibfnamefont {M.}~\bibnamefont {Imada}}, \bibinfo {author} {\bibfnamefont
  {Z.}~\bibnamefont {Gai}}, \bibinfo {author} {\bibfnamefont {M.}~\bibnamefont
  {Pan}}, \ and\ \bibinfo {author} {\bibfnamefont {D.~G.}\ \bibnamefont
  {Mandrus}},\ }\href@noop {} {\bibfield  {journal} {\bibinfo  {journal} {Sci.
  Rep.}\ }\textbf {\bibinfo {volume} {3}},\ \bibinfo {pages} {3073} (\bibinfo
  {year} {2013})}\BibitemShut {NoStop}%
\bibitem [{\citenamefont {Wang}\ and\ \citenamefont
  {Senthil}(2011)}]{FaWang11}%
  \BibitemOpen
  \bibfield  {author} {\bibinfo {author} {\bibfnamefont {F.}~\bibnamefont
  {Wang}}\ and\ \bibinfo {author} {\bibfnamefont {T.}~\bibnamefont {Senthil}},\
  }\href {\doibase 10.1103/PhysRevLett.106.136402} {\bibfield  {journal}
  {\bibinfo  {journal} {Phys. Rev. Lett.}\ }\textbf {\bibinfo {volume} {106}},\
  \bibinfo {pages} {136402} (\bibinfo {year} {2011})}\BibitemShut {NoStop}%
\bibitem [{\citenamefont {Werner}\ and\ \citenamefont
  {Millis}(2006)}]{Werner06}%
  \BibitemOpen
  \bibfield  {author} {\bibinfo {author} {\bibfnamefont {P.}~\bibnamefont
  {Werner}}\ and\ \bibinfo {author} {\bibfnamefont {A.~J.}\ \bibnamefont
  {Millis}},\ }\href {\doibase 10.1103/PhysRevB.74.155107} {\bibfield
  {journal} {\bibinfo  {journal} {Phys. Rev. B}\ }\textbf {\bibinfo {volume}
  {74}},\ \bibinfo {pages} {155107} (\bibinfo {year} {2006})}\BibitemShut
  {NoStop}%
\bibitem [{\citenamefont {Haule}(2007)}]{Haule07}%
  \BibitemOpen
  \bibfield  {author} {\bibinfo {author} {\bibfnamefont {K.}~\bibnamefont
  {Haule}},\ }\href {\doibase 10.1103/PhysRevB.75.155113} {\bibfield  {journal}
  {\bibinfo  {journal} {Phys. Rev. B}\ }\textbf {\bibinfo {volume} {75}},\
  \bibinfo {pages} {155113} (\bibinfo {year} {2007})}\BibitemShut {NoStop}%
\bibitem [{\citenamefont {Huang}\ and\ \citenamefont {Dai}(2012)}]{Huang12a}%
  \BibitemOpen
  \bibfield  {author} {\bibinfo {author} {\bibfnamefont {L.}~\bibnamefont
  {Huang}}\ and\ \bibinfo {author} {\bibfnamefont {X.}~\bibnamefont {Dai}},\
  }\href@noop {} {\bibfield  {journal} {\bibinfo  {journal} {arXiv:1205.1708}\
  } (\bibinfo {year} {2012})}\BibitemShut {NoStop}%
\bibitem [{\citenamefont {Du}\ \emph {et~al.}(2013)\citenamefont {Du},
  \citenamefont {Huang},\ and\ \citenamefont {Dai}}]{Du13b}%
  \BibitemOpen
  \bibfield  {author} {\bibinfo {author} {\bibfnamefont {L.}~\bibnamefont
  {Du}}, \bibinfo {author} {\bibfnamefont {L.}~\bibnamefont {Huang}}, \ and\
  \bibinfo {author} {\bibfnamefont {X.}~\bibnamefont {Dai}},\ }\href {\doibase
  10.1140/epjb/e2013-31024-6} {\bibfield  {journal} {\bibinfo  {journal} {Eur.
  Phys. J. B}\ }\textbf {\bibinfo {volume} {86}},\ \bibinfo {pages} {94}
  (\bibinfo {year} {2013})}\BibitemShut {NoStop}%
\bibitem [{\citenamefont {Dominicis}\ and\ \citenamefont
  {Martin}(1964{\natexlab{a}})}]{Dominicis64a}%
  \BibitemOpen
  \bibfield  {author} {\bibinfo {author} {\bibfnamefont {C.~d.}\ \bibnamefont
  {Dominicis}}\ and\ \bibinfo {author} {\bibfnamefont {P.~C.}\ \bibnamefont
  {Martin}},\ }\href@noop {} {\bibfield  {journal} {\bibinfo  {journal} {J.
  Math. Phys.}\ }\textbf {\bibinfo {volume} {5}},\ \bibinfo {pages} {14}
  (\bibinfo {year} {1964}{\natexlab{a}})}\BibitemShut {NoStop}%
\bibitem [{\citenamefont {Dominicis}\ and\ \citenamefont
  {Martin}(1964{\natexlab{b}})}]{Dominicis64b}%
  \BibitemOpen
  \bibfield  {author} {\bibinfo {author} {\bibfnamefont {C.~d.}\ \bibnamefont
  {Dominicis}}\ and\ \bibinfo {author} {\bibfnamefont {P.~C.}\ \bibnamefont
  {Martin}},\ }\href@noop {} {\bibfield  {journal} {\bibinfo  {journal} {J.
  Math. Phys.}\ }\textbf {\bibinfo {volume} {5}},\ \bibinfo {pages} {31}
  (\bibinfo {year} {1964}{\natexlab{b}})}\BibitemShut {NoStop}%
\bibitem [{\citenamefont {Bickers}\ and\ \citenamefont
  {White}(1991)}]{Bickers91}%
  \BibitemOpen
  \bibfield  {author} {\bibinfo {author} {\bibfnamefont {N.~E.}\ \bibnamefont
  {Bickers}}\ and\ \bibinfo {author} {\bibfnamefont {S.~R.}\ \bibnamefont
  {White}},\ }\href {\doibase 10.1103/PhysRevB.43.8044} {\bibfield  {journal}
  {\bibinfo  {journal} {Phys. Rev. B}\ }\textbf {\bibinfo {volume} {43}},\
  \bibinfo {pages} {8044} (\bibinfo {year} {1991})}\BibitemShut {NoStop}%
\bibitem [{\citenamefont {Yang}\ \emph {et~al.}(2009)\citenamefont {Yang},
  \citenamefont {Fotso}, \citenamefont {Liu}, \citenamefont {Maier},
  \citenamefont {Tomko}, \citenamefont {D'Azevedo}, \citenamefont {Scalettar},
  \citenamefont {Pruschke},\ and\ \citenamefont {Jarrell}}]{SXYang2009}%
  \BibitemOpen
  \bibfield  {author} {\bibinfo {author} {\bibfnamefont {S.~X.}\ \bibnamefont
  {Yang}}, \bibinfo {author} {\bibfnamefont {H.}~\bibnamefont {Fotso}},
  \bibinfo {author} {\bibfnamefont {J.}~\bibnamefont {Liu}}, \bibinfo {author}
  {\bibfnamefont {T.~A.}\ \bibnamefont {Maier}}, \bibinfo {author}
  {\bibfnamefont {K.}~\bibnamefont {Tomko}}, \bibinfo {author} {\bibfnamefont
  {E.~F.}\ \bibnamefont {D'Azevedo}}, \bibinfo {author} {\bibfnamefont {R.~T.}\
  \bibnamefont {Scalettar}}, \bibinfo {author} {\bibfnamefont {T.}~\bibnamefont
  {Pruschke}}, \ and\ \bibinfo {author} {\bibfnamefont {M.}~\bibnamefont
  {Jarrell}},\ }\href {\doibase 10.1103/PhysRevE.80.046706} {\bibfield
  {journal} {\bibinfo  {journal} {Phys. Rev. E}\ }\textbf {\bibinfo {volume}
  {80}},\ \bibinfo {pages} {046706} (\bibinfo {year} {2009})}\BibitemShut
  {NoStop}%
\bibitem [{\citenamefont {Rohringer}\ \emph {et~al.}(2012)\citenamefont
  {Rohringer}, \citenamefont {Valli},\ and\ \citenamefont
  {Toschi}}]{Rohringer12}%
  \BibitemOpen
  \bibfield  {author} {\bibinfo {author} {\bibfnamefont {G.}~\bibnamefont
  {Rohringer}}, \bibinfo {author} {\bibfnamefont {A.}~\bibnamefont {Valli}}, \
  and\ \bibinfo {author} {\bibfnamefont {A.}~\bibnamefont {Toschi}},\ }\href
  {\doibase 10.1103/PhysRevB.86.125114} {\bibfield  {journal} {\bibinfo
  {journal} {Phys. Rev. B}\ }\textbf {\bibinfo {volume} {86}},\ \bibinfo
  {pages} {125114} (\bibinfo {year} {2012})}\BibitemShut {NoStop}%
\bibitem [{\citenamefont {Tam}\ \emph {et~al.}(2013)\citenamefont {Tam},
  \citenamefont {Fotso}, \citenamefont {Yang}, \citenamefont {Lee},
  \citenamefont {Moreno}, \citenamefont {Ramanujam},\ and\ \citenamefont
  {Jarrell}}]{Tam13}%
  \BibitemOpen
  \bibfield  {author} {\bibinfo {author} {\bibfnamefont {K.-M.}\ \bibnamefont
  {Tam}}, \bibinfo {author} {\bibfnamefont {H.}~\bibnamefont {Fotso}}, \bibinfo
  {author} {\bibfnamefont {S.-X.}\ \bibnamefont {Yang}}, \bibinfo {author}
  {\bibfnamefont {T.-W.}\ \bibnamefont {Lee}}, \bibinfo {author} {\bibfnamefont
  {J.}~\bibnamefont {Moreno}}, \bibinfo {author} {\bibfnamefont
  {J.}~\bibnamefont {Ramanujam}}, \ and\ \bibinfo {author} {\bibfnamefont
  {M.}~\bibnamefont {Jarrell}},\ }\href {\doibase 10.1103/PhysRevE.87.013311}
  {\bibfield  {journal} {\bibinfo  {journal} {Phys. Rev. E}\ }\textbf {\bibinfo
  {volume} {87}},\ \bibinfo {pages} {013311} (\bibinfo {year}
  {2013})}\BibitemShut {NoStop}%
\bibitem [{\citenamefont {Maier}\ \emph {et~al.}(2005)\citenamefont {Maier},
  \citenamefont {Jarrell}, \citenamefont {Pruschke},\ and\ \citenamefont
  {Hettler}}]{Maier05}%
  \BibitemOpen
  \bibfield  {author} {\bibinfo {author} {\bibfnamefont {T.}~\bibnamefont
  {Maier}}, \bibinfo {author} {\bibfnamefont {M.}~\bibnamefont {Jarrell}},
  \bibinfo {author} {\bibfnamefont {T.}~\bibnamefont {Pruschke}}, \ and\
  \bibinfo {author} {\bibfnamefont {M.~H.}\ \bibnamefont {Hettler}},\ }\href
  {\doibase 10.1103/RevModPhys.77.1027} {\bibfield  {journal} {\bibinfo
  {journal} {Rev. Mod. Phys.}\ }\textbf {\bibinfo {volume} {77}},\ \bibinfo
  {pages} {1027} (\bibinfo {year} {2005})}\BibitemShut {NoStop}%
\bibitem [{\citenamefont {Gull}\ \emph {et~al.}(2011)\citenamefont {Gull},
  \citenamefont {Millis}, \citenamefont {Lichtenstein}, \citenamefont
  {Rubtsov}, \citenamefont {Troyer},\ and\ \citenamefont {Werner}}]{Gull11}%
  \BibitemOpen
  \bibfield  {author} {\bibinfo {author} {\bibfnamefont {E.}~\bibnamefont
  {Gull}}, \bibinfo {author} {\bibfnamefont {A.~J.}\ \bibnamefont {Millis}},
  \bibinfo {author} {\bibfnamefont {A.~I.}\ \bibnamefont {Lichtenstein}},
  \bibinfo {author} {\bibfnamefont {A.~N.}\ \bibnamefont {Rubtsov}}, \bibinfo
  {author} {\bibfnamefont {M.}~\bibnamefont {Troyer}}, \ and\ \bibinfo {author}
  {\bibfnamefont {P.}~\bibnamefont {Werner}},\ }\href {\doibase
  10.1103/RevModPhys.83.349} {\bibfield  {journal} {\bibinfo  {journal} {Rev.
  Mod. Phys.}\ }\textbf {\bibinfo {volume} {83}},\ \bibinfo {pages} {349}
  (\bibinfo {year} {2011})}\BibitemShut {NoStop}%
\bibitem [{\citenamefont {Chen}\ \emph
  {et~al.}(2013{\natexlab{a}})\citenamefont {Chen}, \citenamefont {Meng},
  \citenamefont {Yu}, \citenamefont {Yang}, \citenamefont {Jarrell},\ and\
  \citenamefont {Moreno}}]{Chen13a}%
  \BibitemOpen
  \bibfield  {author} {\bibinfo {author} {\bibfnamefont {K.~S.}\ \bibnamefont
  {Chen}}, \bibinfo {author} {\bibfnamefont {Z.~Y.}\ \bibnamefont {Meng}},
  \bibinfo {author} {\bibfnamefont {U.}~\bibnamefont {Yu}}, \bibinfo {author}
  {\bibfnamefont {S.}~\bibnamefont {Yang}}, \bibinfo {author} {\bibfnamefont
  {M.}~\bibnamefont {Jarrell}}, \ and\ \bibinfo {author} {\bibfnamefont
  {J.}~\bibnamefont {Moreno}},\ }\href {\doibase 10.1103/PhysRevB.88.041103}
  {\bibfield  {journal} {\bibinfo  {journal} {Phys. Rev. B}\ }\textbf {\bibinfo
  {volume} {88}},\ \bibinfo {pages} {041103} (\bibinfo {year}
  {2013}{\natexlab{a}})}\BibitemShut {NoStop}%
\bibitem [{\citenamefont {Chen}\ \emph
  {et~al.}(2013{\natexlab{b}})\citenamefont {Chen}, \citenamefont {Meng},
  \citenamefont {Yang}, \citenamefont {Pruschke}, \citenamefont {Moreno},\ and\
  \citenamefont {Jarrell}}]{Chen13b}%
  \BibitemOpen
  \bibfield  {author} {\bibinfo {author} {\bibfnamefont {K.-S.}\ \bibnamefont
  {Chen}}, \bibinfo {author} {\bibfnamefont {Z.~Y.}\ \bibnamefont {Meng}},
  \bibinfo {author} {\bibfnamefont {S.-X.}\ \bibnamefont {Yang}}, \bibinfo
  {author} {\bibfnamefont {T.}~\bibnamefont {Pruschke}}, \bibinfo {author}
  {\bibfnamefont {J.}~\bibnamefont {Moreno}}, \ and\ \bibinfo {author}
  {\bibfnamefont {M.}~\bibnamefont {Jarrell}},\ }\href {\doibase
  10.1103/PhysRevB.88.245110} {\bibfield  {journal} {\bibinfo  {journal} {Phys.
  Rev. B}\ }\textbf {\bibinfo {volume} {88}},\ \bibinfo {pages} {245110}
  (\bibinfo {year} {2013}{\natexlab{b}})}\BibitemShut {NoStop}%
\bibitem [{\citenamefont {Arita}\ \emph {et~al.}(2012)\citenamefont {Arita},
  \citenamefont {Kunes}, \citenamefont {Kozhevnikov}, \citenamefont {Eguiluz},\
  and\ \citenamefont {Imada}}]{Arita12}%
  \BibitemOpen
  \bibfield  {author} {\bibinfo {author} {\bibfnamefont {R.}~\bibnamefont
  {Arita}}, \bibinfo {author} {\bibfnamefont {J.}~\bibnamefont {Kunes}},
  \bibinfo {author} {\bibfnamefont {A.~V.}\ \bibnamefont {Kozhevnikov}},
  \bibinfo {author} {\bibfnamefont {A.~G.}\ \bibnamefont {Eguiluz}}, \ and\
  \bibinfo {author} {\bibfnamefont {M.}~\bibnamefont {Imada}},\ }\href
  {\doibase 10.1103/PhysRevLett.108.086403} {\bibfield  {journal} {\bibinfo
  {journal} {Phys. Rev. Lett.}\ }\textbf {\bibinfo {volume} {108}},\ \bibinfo
  {pages} {086403} (\bibinfo {year} {2012})}\BibitemShut {NoStop}%
\bibitem [{\citenamefont {Watanabe}\ \emph {et~al.}(2013)\citenamefont
  {Watanabe}, \citenamefont {Shirakawa},\ and\ \citenamefont
  {Yunoki}}]{Watanabe13}%
  \BibitemOpen
  \bibfield  {author} {\bibinfo {author} {\bibfnamefont {H.}~\bibnamefont
  {Watanabe}}, \bibinfo {author} {\bibfnamefont {T.}~\bibnamefont {Shirakawa}},
  \ and\ \bibinfo {author} {\bibfnamefont {S.}~\bibnamefont {Yunoki}},\ }\href
  {\doibase 10.1103/PhysRevLett.110.027002} {\bibfield  {journal} {\bibinfo
  {journal} {Phys. Rev. Lett.}\ }\textbf {\bibinfo {volume} {110}},\ \bibinfo
  {pages} {027002} (\bibinfo {year} {2013})}\BibitemShut {NoStop}%
\bibitem [{\citenamefont {Yang}\ \emph {et~al.}(2014)\citenamefont {Yang},
  \citenamefont {Wang}, \citenamefont {Liu}, \citenamefont {Chen},
  \citenamefont {Dai},\ and\ \citenamefont {Wang}}]{Wang14}%
  \BibitemOpen
  \bibfield  {author} {\bibinfo {author} {\bibfnamefont {Y.}~\bibnamefont
  {Yang}}, \bibinfo {author} {\bibfnamefont {W.-S.}\ \bibnamefont {Wang}},
  \bibinfo {author} {\bibfnamefont {J.-G.}\ \bibnamefont {Liu}}, \bibinfo
  {author} {\bibfnamefont {H.}~\bibnamefont {Chen}}, \bibinfo {author}
  {\bibfnamefont {J.-H.}\ \bibnamefont {Dai}}, \ and\ \bibinfo {author}
  {\bibfnamefont {Q.-H.}\ \bibnamefont {Wang}},\ }\href {\doibase
  10.1103/PhysRevB.89.094518} {\bibfield  {journal} {\bibinfo  {journal} {Phys.
  Rev. B}\ }\textbf {\bibinfo {volume} {89}},\ \bibinfo {pages} {094518}
  (\bibinfo {year} {2014})}\BibitemShut {NoStop}%
\bibitem [{\citenamefont {Han}(2004)}]{Han04}%
  \BibitemOpen
  \bibfield  {author} {\bibinfo {author} {\bibfnamefont {J.~E.}\ \bibnamefont
  {Han}},\ }\href {\doibase 10.1103/PhysRevB.70.054513} {\bibfield  {journal}
  {\bibinfo  {journal} {Phys. Rev. B}\ }\textbf {\bibinfo {volume} {70}},\
  \bibinfo {pages} {054513} (\bibinfo {year} {2004})}\BibitemShut {NoStop}%
\bibitem [{\citenamefont {Sakai}\ \emph {et~al.}(2004)\citenamefont {Sakai},
  \citenamefont {Arita},\ and\ \citenamefont {Aoki}}]{Sakai04}%
  \BibitemOpen
  \bibfield  {author} {\bibinfo {author} {\bibfnamefont {S.}~\bibnamefont
  {Sakai}}, \bibinfo {author} {\bibfnamefont {R.}~\bibnamefont {Arita}}, \ and\
  \bibinfo {author} {\bibfnamefont {H.}~\bibnamefont {Aoki}},\ }\href {\doibase
  10.1103/PhysRevB.70.172504} {\bibfield  {journal} {\bibinfo  {journal} {Phys.
  Rev. B}\ }\textbf {\bibinfo {volume} {70}},\ \bibinfo {pages} {172504}
  (\bibinfo {year} {2004})}\BibitemShut {NoStop}%
\bibitem [{\citenamefont {Puetter}\ and\ \citenamefont
  {Kee}(2012)}]{Puetter12}%
  \BibitemOpen
  \bibfield  {author} {\bibinfo {author} {\bibfnamefont {C.~M.}\ \bibnamefont
  {Puetter}}\ and\ \bibinfo {author} {\bibfnamefont {H.-Y.}\ \bibnamefont
  {Kee}},\ }\href@noop {} {\bibfield  {journal} {\bibinfo  {journal} {EPL}\
  }\textbf {\bibinfo {volume} {98}},\ \bibinfo {pages} {27010} (\bibinfo {year}
  {2012})}\BibitemShut {NoStop}%
\bibitem [{sup()}]{supplemental}%
  \BibitemOpen
  \href@noop {} {\bibinfo  {journal} {See Supplemental Material at
  http://link.aps.org/supplemental/10.1103/PhysRevLett.113.177003, which
  includes
  Refs.~\onlinecite{Cao98,BJKim08,BJKim09,Fujiyama12,Carter12,Carter13,Watanabe10,Watanabe13,Watanabe14,Wang14,Du13b,Bickers91,SXYang2009,Rohringer12,Tam13},
  for details about the t2g three-orbital model, the CTQMC/DMFT+Parquet
  formalism and its numerical implementation and performance.}\ }\BibitemShut
  {NoStop}%
\bibitem [{\citenamefont {Watanabe}\ \emph {et~al.}(2010)\citenamefont
  {Watanabe}, \citenamefont {Shirakawa},\ and\ \citenamefont
  {Yunoki}}]{Watanabe10}%
  \BibitemOpen
\bibfield  {journal} {  }\bibfield  {author} {\bibinfo {author} {\bibfnamefont
  {H.}~\bibnamefont {Watanabe}}, \bibinfo {author} {\bibfnamefont
  {T.}~\bibnamefont {Shirakawa}}, \ and\ \bibinfo {author} {\bibfnamefont
  {S.}~\bibnamefont {Yunoki}},\ }\href {\doibase
  10.1103/PhysRevLett.105.216410} {\bibfield  {journal} {\bibinfo  {journal}
  {Phys. Rev. Lett.}\ }\textbf {\bibinfo {volume} {105}},\ \bibinfo {pages}
  {216410} (\bibinfo {year} {2010})}\BibitemShut {NoStop}%
\bibitem [{\citenamefont {Watanabe}\ \emph {et~al.}(2014)\citenamefont
  {Watanabe}, \citenamefont {Shirakawa},\ and\ \citenamefont
  {Yunoki}}]{Watanabe14}%
  \BibitemOpen
  \bibfield  {author} {\bibinfo {author} {\bibfnamefont {H.}~\bibnamefont
  {Watanabe}}, \bibinfo {author} {\bibfnamefont {T.}~\bibnamefont {Shirakawa}},
  \ and\ \bibinfo {author} {\bibfnamefont {S.}~\bibnamefont {Yunoki}},\ }\href
  {\doibase 10.1103/PhysRevB.89.165115} {\bibfield  {journal} {\bibinfo
  {journal} {Phys. Rev. B}\ }\textbf {\bibinfo {volume} {89}},\ \bibinfo
  {pages} {165115} (\bibinfo {year} {2014})}\BibitemShut {NoStop}%
\bibitem [{foo()}]{footnote14}%
  \BibitemOpen
  \href@noop {} {\bibinfo  {journal} {Similar interaction parameters are used
  in Ref.~\cite{Arita12,Watanabe13}, but phases at finite $J$ in the hole-doped
  side are not studied}\ }\BibitemShut {NoStop}%
\bibitem [{\citenamefont {Schrieffer}(1983)}]{Schrieffer83}%
  \BibitemOpen
\bibfield  {journal} {  }\bibfield  {author} {\bibinfo {author} {\bibfnamefont
  {J.~R.}\ \bibnamefont {Schrieffer}},\ }\href@noop {} {\emph {\bibinfo {title}
  {Theory of Superconductivity}}}\ (\bibinfo  {publisher} {ABC, Westview
  Press},\ \bibinfo {year} {1983})\BibitemShut {NoStop}%
\bibitem [{\citenamefont {Sigrist}\ and\ \citenamefont
  {Ueda}(1991)}]{Sigrist91}%
  \BibitemOpen
  \bibfield  {author} {\bibinfo {author} {\bibfnamefont {M.}~\bibnamefont
  {Sigrist}}\ and\ \bibinfo {author} {\bibfnamefont {K.}~\bibnamefont {Ueda}},\
  }\href {\doibase 10.1103/RevModPhys.63.239} {\bibfield  {journal} {\bibinfo
  {journal} {Rev. Mod. Phys.}\ }\textbf {\bibinfo {volume} {63}},\ \bibinfo
  {pages} {239} (\bibinfo {year} {1991})}\BibitemShut {NoStop}%
\bibitem [{\citenamefont {Korneta}\ \emph {et~al.}(2010)\citenamefont
  {Korneta}, \citenamefont {Qi}, \citenamefont {Chikara}, \citenamefont
  {Parkin}, \citenamefont {De~Long}, \citenamefont {Schlottmann},\ and\
  \citenamefont {Cao}}]{Korneta10}%
  \BibitemOpen
  \bibfield  {author} {\bibinfo {author} {\bibfnamefont {O.~B.}\ \bibnamefont
  {Korneta}}, \bibinfo {author} {\bibfnamefont {T.}~\bibnamefont {Qi}},
  \bibinfo {author} {\bibfnamefont {S.}~\bibnamefont {Chikara}}, \bibinfo
  {author} {\bibfnamefont {S.}~\bibnamefont {Parkin}}, \bibinfo {author}
  {\bibfnamefont {L.~E.}\ \bibnamefont {De~Long}}, \bibinfo {author}
  {\bibfnamefont {P.}~\bibnamefont {Schlottmann}}, \ and\ \bibinfo {author}
  {\bibfnamefont {G.}~\bibnamefont {Cao}},\ }\href {\doibase
  10.1103/PhysRevB.82.115117} {\bibfield  {journal} {\bibinfo  {journal} {Phys.
  Rev. B}\ }\textbf {\bibinfo {volume} {82}},\ \bibinfo {pages} {115117}
  (\bibinfo {year} {2010})}\BibitemShut {NoStop}%
\bibitem [{\citenamefont {Huang}\ \emph {et~al.}(2014)\citenamefont {Huang},
  \citenamefont {Wang}, \citenamefont {Meng}, \citenamefont {Du}, \citenamefont
  {Werner},\ and\ \citenamefont {Dai}}]{iQIST14}%
  \BibitemOpen
  \bibfield  {author} {\bibinfo {author} {\bibfnamefont {L.}~\bibnamefont
  {Huang}}, \bibinfo {author} {\bibfnamefont {Y.}~\bibnamefont {Wang}},
  \bibinfo {author} {\bibfnamefont {Z.~Y.}\ \bibnamefont {Meng}}, \bibinfo
  {author} {\bibfnamefont {L.}~\bibnamefont {Du}}, \bibinfo {author}
  {\bibfnamefont {P.}~\bibnamefont {Werner}}, \ and\ \bibinfo {author}
  {\bibfnamefont {X.}~\bibnamefont {Dai}},\ }\href@noop {} {\bibfield
  {journal} {\bibinfo  {journal} {arXiv:1409.7573}\ } (\bibinfo {year}
  {2014})}\BibitemShut {NoStop}%
\end{thebibliography}%

\end{document}